\documentclass[aps,prl,preprint,showpacs,preprintnumbers,amsmath,amssymb]{revtex4}

\usepackage{graphicx}
\usepackage{dcolumn}
\usepackage{bm}

\begin{document}

\title{Partial quantum statistics and its implications for narrow band materials}

\author{T. Zhou}

\email{tzhou2@lucent.com}

\affiliation{Bell Laboratories, Lucent Technologies, 791 Holmdel
Road, Holmdel, New Jersey 07733, U.S.A.}


\begin{abstract}
Based upon the newly proposed partial quantum statistics [T. Zhou,
Solid State Commun. {\bf 115}, 185 (2000)], some canonical
physical properties of partially localized electron systems have
been calculated. The calculated transport and superconducting
properties of such systems are very different from those of
Landau-Fermi liquids, but display some striking similarities to
the properties of high temperature superconductors and some other
narrow band materials. \\

\end{abstract}


\maketitle

Bloch states are essentially unbound states. A Bloch wave function
at any given time is nonzero in the whole crystal lattice except
its nodal points. Because of their identical intrinsic properties
and delocalized nature Bloch electrons are indistinguishable from
each other. Neglecting their mutual interactions, a system
consisting of Bloch electrons is thus an ideal Fermi gas system
and strictly obeys the Fermi-Dirac (F.D.) statistics at any
temperature, though at high temperature the Maxwell-Boltzmann
(M.B.) distribution is a good approximation. On the other end of
the spectrum, when electrons are localized and completely
separated from each other like the electrons of magnetic ions in
paramagnetic salts, these electrons are in bound states, and the
wave function of one electron (or the many-body wave function of
several electrons of the same ion) does not overlap with another
spatially. These electrons thus become distinguishable, and the
system strictly obeys the M.B. distribution at any temperature
\cite{wann66}. This is of course in consistent with the Langevin
paramagnetic behavior observed in paramagnetic salts down to very
low temperature \cite{wann66,kitt71}, and in contrast to the Pauli
paramagnetism expected in Fermi gas systems. Bloch electrons, on
the other hand, can be localized in real space by strong
electron-electron interaction (large Hubbard U), strong
electron-phonon interaction ({\em e.g.} small polaron), and/or by
disorder (Anderson localization). In fact, paramagnetic salts with
ions containing only one 3d or 4f electron (or hole) can be taken
as an extreme and simple example where the Hubbard U is infinite
and no site can be occupied by two electrons (or two holes)
simultaneously. Now the question is, when the electrons are
neither completely localized as in the paramagnetic salts, nor
completely delocalized as the Bloch electrons, what kind of
statistical distribution should they obey? Should the electrons,
after their mutual interactions are taken into account by
renormalization, always adopt the F.D. statistics as indicated in
the Landau Fermi-liquid theorem?

In a gedanken experiment recently proposed by the author
\cite{zhou00}, it has been shown that in any self-consistent
theory for partially localized electrons (PLE), neither the F.D.
nor the M.B. distribution can be the right answer, and there has
to be a partial Fermi (p.F.) statistics which is different from
both the F.D. and M.B. statistics. Furthermore, it has been argued
by the author that in order to reach such a p.F. statistics, the
Slater-determinant type many-body wave function used to describe
the Fermi gas system has to be modified, and the many-body wave
function proposed by the author \cite{zhou00} violates the
anti-symmetry requirement for the wave function of identical
fermions. The particle exchange symmetry thus becomes a broken
symmetry, and it was argued in Ref. \cite{zhou00} that the
breaking of such a symmetry in PLE systems does not violate any
fundamental quantum mechanics principle, including Pauli's
exclusion principle. The conventional belief that quantum field
theory requires the wave function of identical fermions to be
anti-symmetric is not applicable here. The reason is that quantum
field theory always assumes, explicitly or implicitly, that
identical particles are indistinguishable. This assumption is not
true here because localized electrons can be distinguished. Based
on this new form of wave function a parameter $\eta$, which is a
real number between 0 and 1, is defined to describe the
indistinguishable degree of PLE system \cite{zhou00}. When $\eta =
1$, the electrons are completely delocalized and
indistinguishable; when $\eta = 0$, the electrons are completely
localized and distinguishable. The p.F. statistical distribution
is subsequently deduced to describe the PLE gas system. Suppose
$f_{pF}(E_{l}, T)$ is the probable number of electrons with
indistinguishable degree $\eta$ occupying any state with an energy
level $E_{l}$ at temperature $T$, one then has \cite{zhou00}
\begin{eqnarray}
&&f_{pF}(E_{l},T) = \frac{1}{e^{(E_{l}-\mu)/k_{B}T} + \eta},
\nonumber\\
&&\sum_{l} \omega_{l} f_{pF}(E_{l},T) = N .
\label{pf}
\end{eqnarray}
Here $k_{B}$ is the Boltzmann constant, $\omega_{l}$ is the
degeneracy degree of the $l$th energy level, and the chemical
potential $\mu$ is still determined by the total electron number
$N$. It is evident from Eq. (\ref{pf}) that when $\eta$ = 1 or 0,
the F.D. or M.B. distribution is recovered, respectively
\cite{zhou00}. We note that the exact mathematical form of Eq.
(\ref{pf}) has appeared in Ref. \cite{acha94} and \cite{poly96},
but there the physical contexts were very different.

It was proposed \cite{zhou00} that $f_{pF}$ should have
significant deviation from the F.D. or M.B distribution in
narrow-band materials \cite{mott90}, since electrons are neither
very delocalized nor very localized in these materials. In this
letter, we will demonstrate that systems obeying the p.F.
statistics have properties that are very different from those of
Landau-Fermi liquids, but are strikingly similar to many
properties of high temperature superconductors and some other
narrow band materials.

We first examine Eq. (\ref{pf}), in which the summation of all
states can be replaced by the integral $\int_{0}^{\infty}
g(E)f_{pF}(E,T)dE = N$. Here $g(E)$ is the total density of
states, which includes both the delocalized and localized
components. $g(E)$ for a p.F. gas system is the same as a Fermi
gas system with the same energy spectrum. The effective
delocalization density of states $g'(E)$ in a p.F. system with
electron indistinguishable degree $\eta$, however, is only
\begin{eqnarray}
g'(E) = \eta g(E),
\label{ge}
\end{eqnarray}
according to the definition of $\eta$ \cite{zhou00}. Replacing
$g(E)$ by $g'(E)$, and rewriting Eq. (\ref{pf}) so that it is
easier to compare it with the F.D. distribution, one reaches
\begin{eqnarray}
&&f'_{pF}(E,T) =
\frac{\eta^{-1}}{e^{[E-(E_{pF}+k_{B}T\ln{\eta})]/k_{B}T} + 1},
\nonumber \\
&&\int_{0}^{\infty} g'(E)f'_{pF}(E,T)dE = N.
\label{pf2}
\end{eqnarray}
Here $f'_{pF}(E,T)$ is the distribution function for the
delocalization component of a p.F. gas system. The chemical
potential $\mu$ is replaced by $E_{pF}$, the effective Fermi
energy in a p.F. system for the delocalization component, and we
call it partial Fermi energy. Assuming that a Fermi gas system
with Fermi energy $E_{F}$ has the same $g(E)$ as the p.F. gas
system in discussion, and $E_{F}^{0}$ is the Fermi energy at $T =
0$, solving Eq. (\ref{pf2}) one has $E_{pF}(T=0)=E_{F}^{0}$, and
$f'_{pF}(T=0) = \eta^{-1}$ or 0 for $E <$ or $ > E_{F}^{0}$,
respectively. At $T > 0$ but $k_{B}T \ll E_{F}^{0}$, solving Eq.
(\ref{pf2}) again and we find that to the first order
approximation of $T$,

\begin{eqnarray}
E_{pF}(T) \approx E_{F}^{0} + k_{B}T \ln \eta^{-1}.
\label{epf}
\end{eqnarray}
Here we have assumed that $g(E)$ is a smooth function near
$E^{0}_{F}$. Combining Eqs. (\ref{pf2}) and (\ref{epf}) one
reaches
\begin{eqnarray}
f'_{pF}(E,T)\approx \frac{\eta^{-1}}{e^{(E-E_{F}^{0})/k_{B}T}+1}
\approx \eta^{-1}f_{F}(E,T). \label{pf3}
\end{eqnarray}
Here $f_{F}(E,T)$ is the F.D. distribution function. Approximating
to the first order of $T$, $E_{F}(T) \approx E_{F}^{0}$ in a Fermi
gas system \cite{kitt71}. We also note that in a p.F. system, if
the spin degree of freedom can be ignored, then $g'(E)$, rather
than $g(E)$, is the density of states that is mostly relevant to
the observable physical properties. The schematic illustration of
the p.F. and F.D. distributions at finite temperature is shown in
Fig. 1.

Equations (\ref{ge}), (\ref{epf}), and (\ref{pf3}) essentially
summarize the differences between the Fermi statistics and the
p.F. statistics at low temperature. In the following we will use
these equations to calculate some canonical thermodynamic,
transport and superconducting properties of p.F. systems. To
compare these calculations with the physical properties of some
real narrow band materials, we take cuprates as the prime examples
\cite{cmms00,mill00}. Undoped cuprates are Mott-Hubbard
insulators, which are extremely localized electron systems. The
overdoped cuprates with doping concentration $x
> 0.3$, however, are believed to be Fermi-liquid systems, and
carriers in these materials are essentially Bloch type and
delocalized. Carriers in cuprates with $0 < x < 0.3$ are believed
to be between these two extreme cases and partially localized.
Cuprates thus provide us a perfect set of real systems, whose
physics is closely associated with the delocalization degree of
the carriers. The discussion above led us to believe that the
delocalization degree of carriers in cuprates can be quantified by
$\eta$, and $\eta$ should increase monotonically with the doping
concentration $x$.

We first calculate the electronic specific heat, one of the most
important thermodynamic properties, of a p.F. gas system. For a
Fermi gas system the electronic specific heat $C_{el}$ at
temperature $k_{B}T \ll E_{F}$ is proportional to $T$ with a
linear coefficient $\gamma = {1 \over 3} \pi^{2}
k_{B}^{2}g(E_{F})$ \cite{kitt71}. For a p.F. gas system, replacing
$g(E)$ and $f_{F}(E,T)$ by Eqs. (\ref{ge}) and (\ref{pf3}),
respectively, one finds that at temperature $k_{B}T \ll E_{pF}$,
$C_{el}$ is the same as the Fermi gas system with the same total
density of states $g(E)$. The reason is that the total energy of a
Fermi or p.F. gas system is decided by $g(E)f_{F}(E,T)$ or
$g'(E)f'_{pF}(E,T)$. Since these two products are equal according
to Eqs. (\ref{ge}) and (\ref{pf3}), $C_{el}$, the derivative of
the total energy, should also be equal in these two systems.
Experiments show that for optimally doped and overdoped cuprates,
the normal state $\gamma$ is indeed independent of temperature and
doping concentration at least up to 300 K \cite{lora94}. This is
consistent with our calculation above of the p.F. gas system. In
the underdoped regime, below a characteristic temperature $T^{*}$
$\gamma$ decreases with decreasing $T$, and this is believed to be
related to the pseudogap \cite{lora94}. Since our discussion above
does not take into account any anomaly in $g'(E)$ such as a
pseudogap, our calculation is thus not comparable with the
experimental results in the underdoped cuprates. Other
thermodynamic properties that are also decided by
$g'(E)f'_{pF}(E,T)$ should yield similar results as $C_{el}$, and
experiments show that the magnetic susceptibility data are indeed
consistent with the $C_{el}$ results in the optimally doped to
overdoped regime \cite{lora94}.

We now turn to the calculation of transport properties. For
Fermi-liquid metals, the dominant scattering mechanism at low
temperature is electron-electron scattering, in which the total
energy and momentum should be conserved while the Fermi statistics
should also be obeyed. The Fermi statistics acts twice, each time
reduces the scattering rate $\tau^{-1}_{ee}$ approximately by a
factor of $k_{B}T/E_{F}$, or $\hbar\omega /E_{F}$ in the case of
zero temperature low energy excitation with excitation energy
$\hbar\omega$ \cite{kitt71}. Suppose $\tau^{-1}_{cee}$ is the
classical electron-electron scattering rate after taking into
account the Coulomb screening effect but without the consideration
of the Fermi statistics. The scattering rate after including the
Fermi statistics then becomes $\tau^{-1}_{ee} \approx
(k_{B}T/E_{F})^{2}\tau^{-1}_{cee}$ or
$(\hbar\omega/E_{F})^{2}\tau^{-1}_{cee}$ \cite{kitt71}. To
calculate $\tau^{-1}_{ee}$ in a p.F. system, one should replace
the Fermi statistics by the p.F. statistics using Eq. (\ref{pf3}).
($\tau^{-1}_{ee}$ as a function of $T$ or $\omega$ is not
dependent on the density of states). Equation (\ref{pf3}), on the
other hand, indicates that the restriction on the
electron-electron scattering process imposed by the Fermi
statistics is partially lifted in a p.F. system, with the
parameter $\eta^{-1}$ linearly interpolating between the Fermi
statistics and the purely classical M.B. statistics where there is
no statistical restriction at all. Therefore, one can express
$\tau^{-1}_{ee}$ of a p.F. system as
\begin{eqnarray}
\tau^{-1}_{ee} & \approx &
(k_{B}T/E^{0}_{F})^{2\eta}\tau^{-1}_{cee}
 \ \ \ \ or \nonumber \\
\tau^{-1}_{ee} & \approx &
(\hbar\omega/E^{0}_{F})^{2\eta}\tau^{-1}_{cee}.
\label{tau}
\end{eqnarray}
Note that $E_{F}$ is replaced by $E^{0}_{F}$ here due to Eq.
(\ref{pf3}). For $\eta$ = 1 or 0, the Fermi-liquid behavior or the
classical behavior is recovered, respectively.

When $\eta$ = ${1 \over 2}$, one has $\tau^{-1}_{ee} \sim T \ or \
\omega$, which is the well known marginal Fermi-liquid (MFL)
phenomenology formulated to describe optimally doped cuprates
\cite{varm89,litt91}. As analyzed by the MFL theory
\cite{varm89,litt91,varm97}, this linear $T$ or $\omega$
dependence explains many transport properties of optimally doped
cuprates, including an electronic Raman background which is both
$T$ and $\omega$ independent, an optical conductivity which
decreases with $\omega$ like $\omega^{-1}$, a dc resistivity
$\rho$ which has the celebrated linear $T$ dependence, {\it etc.}
With $x$ increasing from the optimally doping value, $\eta$ is
expected to increase from ${1 \over 2}$ to 1, as discussed above.
Equation (\ref{tau}) thus also naturally explains the gradual
crossover from the MFL behavior to Fermi-liquid behavior with
increasing doping, which is best evidenced experimentally by
$\rho$'s gradual change from $T$ to $T^{2}$ dependence with
increasing doping in the optimally doped to overdoped regime
\cite{batl94,batl98}. Furthermore, $\rho$ of underdoped cuprates
in the intermediate $T$ range above the temperature of the
insulating regime may be fitted by $T^{\alpha}$ (see the data in
Ref. \cite{batl94,batl98}), with $\alpha$ smaller than 1 and
decreasing with decreasing $x$. This is consistent with Eq.
(\ref{pf3}), where $\eta$ is expected to be smaller than ${1 \over
2}$ and decrease with $x$ in the underdoped regime. Since $k_{B}T
\ll E^{0}_{F}$ for $T < 300 K$, $\tau^{-1}_{ee}$ should increase
significantly with decreasing $\eta$, according to Eq.
(\ref{tau}). This means that in the optimally doped to underdoped
regime, at room temperature or even above, the electron-electron
scattering can still be dominant over other scattering mechanisms,
including the electron-phonon scattering. This helps to explain
why the linear $T$ dependence of $\rho$ can be seen up to very
high temperature \cite{mart90}. In addition, a low temperature
$\rho \sim T^{\alpha}$ behavior with $0 < \alpha < 2$ has been
observed in other narrow band materials, two of the most recent
examples are BaVS$_{3}$ \cite{forr00} and La$_{4}$Ru$_{6}$O$_{19}$
\cite{khal01}. In BaVS$_{3}$ $\alpha$ can even be tuned by
pressure \cite{forr00}. These are all consistent with Eq.
(\ref{tau}) predicted by the p.F. statistics.

We finally come to the calculation of the zero temperature
superconducting gap $\Delta(0)$ and the superconducting critical
temperature $T_{c}$ of a p.F. system. A p.F. gas system of course
can not superconduct, so we first introduce a BCS-type attractive
interaction $V$. Similar to the assumption in the BCS theory
\cite{bard57}, $V$ is supposed to be constant between two
partially localized electrons with energies immediately below
$E_{pF}$ within a shell of $\hbar\omega_{D}$ thick, and zero for
any other electron pairs. $\hbar\omega_{D}$ is the energy of
phonons, or any other types of bosons that mediate the formation
of Cooper pairs. Using the same variational method as in the
deduction of the BCS theory \cite{bard57}, and replacing the Fermi
statistics by Eqs. (\ref{ge}), (\ref{epf}), and (\ref{pf3}), one
reaches
\begin{eqnarray}
\Delta(0) & \approx & 2\hbar\omega_{D} \ e^{-\eta/g(E_{F}^{0})V},
\nonumber \\
k_{B}T_{c} & \approx & \hbar\omega_{D}[0.783 \ e^{2\eta /
g(E_{F}^{0})V} + (\ln \eta)^{2}]^{-1/2}.
\label {tc}
\end{eqnarray}
When $\eta = 1$, the BCS expression of $\Delta(0)$ and $T_{c}$ are
recovered. When $\eta = 0$, $\Delta(0) = k_{B}T_{c} = 0$. For
$g(E_{F}^{0})V = 0.2$, which is a typical weak coupling value, the
maximum of $T_{c}$ (denoted as $T_{cm}$) is about
$0.38\hbar\omega_{D}$ with a corresponding $\eta \approx 0.135$,
as shown in Fig. 2. This is about 50 times of the BCS $T_{c}$
value with the same $g(E_{F}^{0})V$. With increasing
$g(E_{F}^{0})V$, $T_{cm}$ increases, so does the corresponding
value of $\eta$ (denoted as $\eta_{m}$). With $g(E_{F}^{0})V =
1.3$, $\eta_{m}$ is about 0.5, the value which we assigned to
optimally doped cuprates in the transport properties discussion
above. However, Fig. 2 should not be used to compare directly with
$T_{c}$ {\it vs.} $x$ of superconducting cuprates, since cuprates
are d-wave superconductors, and the assumption of $V$ above cannot
be applied. Moreover, even for phonon mediating s-wave
superconductors a direct comparison is also not possible, because
in order to have an $\eta$ significantly deviating from 1, the
electron-phonon interaction will be too strong to allow the
assumption of a weak-coupling $V$. Nevertheless, we believe that
no matter what specific form $V$ takes, the basic feature of
$T_{c}$ {\it vs.} $\eta$ should remain the same for systems
obeying the p.F. statistics. That is, with increasing $\eta$,
$T_{c}$ should always first increase from 0, reach a maximum, then
decrease monotonically until $\eta = 1$. The reason is that this
feature is decided by the p.F. statistics, not by the specific
form of $V$. In fact, the well known $T_{c} \ vs. \ n_{s}/m^{*}$
plot produced by Uemura {\it et al.} \cite{uemu91} is
qualitatively similar to the plots in Fig. 2. Note that $n_{s}$ is
the density of superconducting carriers, $m^{*}$ is the effective
mass. Since $\eta$ is the delocalization degree of p.F. systems,
it is thus reasonable to assume that $\eta$ increases
monotonically with $n_{s}/m^{*}$. We also note that the Uemura
plot includes many different systems with different
superconducting mechanisms, but all of them are narrow band
materials \cite{uemu91,uemu91b}. Recently spectacular results have
been achieved on $C_{60}$ with injected carriers, and $T_{c}$ has
reached 117 K \cite{scho00}. In these materials, $T_{c}$ as a
function of injected carrier concentration again shows the same
basic feature mentioned above \cite{scho00}. Because of the narrow
band width of $C_{60}$ crystals \cite{uemu91b}, increasing
injected carriers may also have increased the delocalization
degree of the carriers in $C_{60}$, and subsequently caused the
change in $T_{c}$.


There are still many open questions for the p.F. systems. For
example, if the spin degree of freedom cannot be ignored, as in
the case of underdoped cuprates which are very close to the
antiferromagnetic phase, what kind of change will it bring to the
p.F. systems? If the localized electrons become mobile with
increasing temperature, how does the partial Fermi energy respond
to such a change, especially at high temperature? If electrons are
more localized in one direction than another, $\eta$ might become
anisotropic, and how is the anisotropy of $\eta$ in k-space
dependent on the many-body wave function of partially localized
electrons? Despite these open questions, this letter nevertheless
has demonstrated that materials obeying the p.F. statistics have
many properties that are very different from Landau-Fermi liquids,
but are consistent with those observed in cuprates and other
narrow band materials. We thus believe that the p.F. statistics
can not only help to understand why bound electron systems are so
different from unbound electron systems \cite{ande00}, it may also
provide a starting point to tackle the physics of cuprates and
other non-Fermi-liquid metals.

\begin{acknowledgments}
I am very thankful to Dr. D.T. Neilson for his support.
\end{acknowledgments}

\clearpage

\begin{figure}
\includegraphics{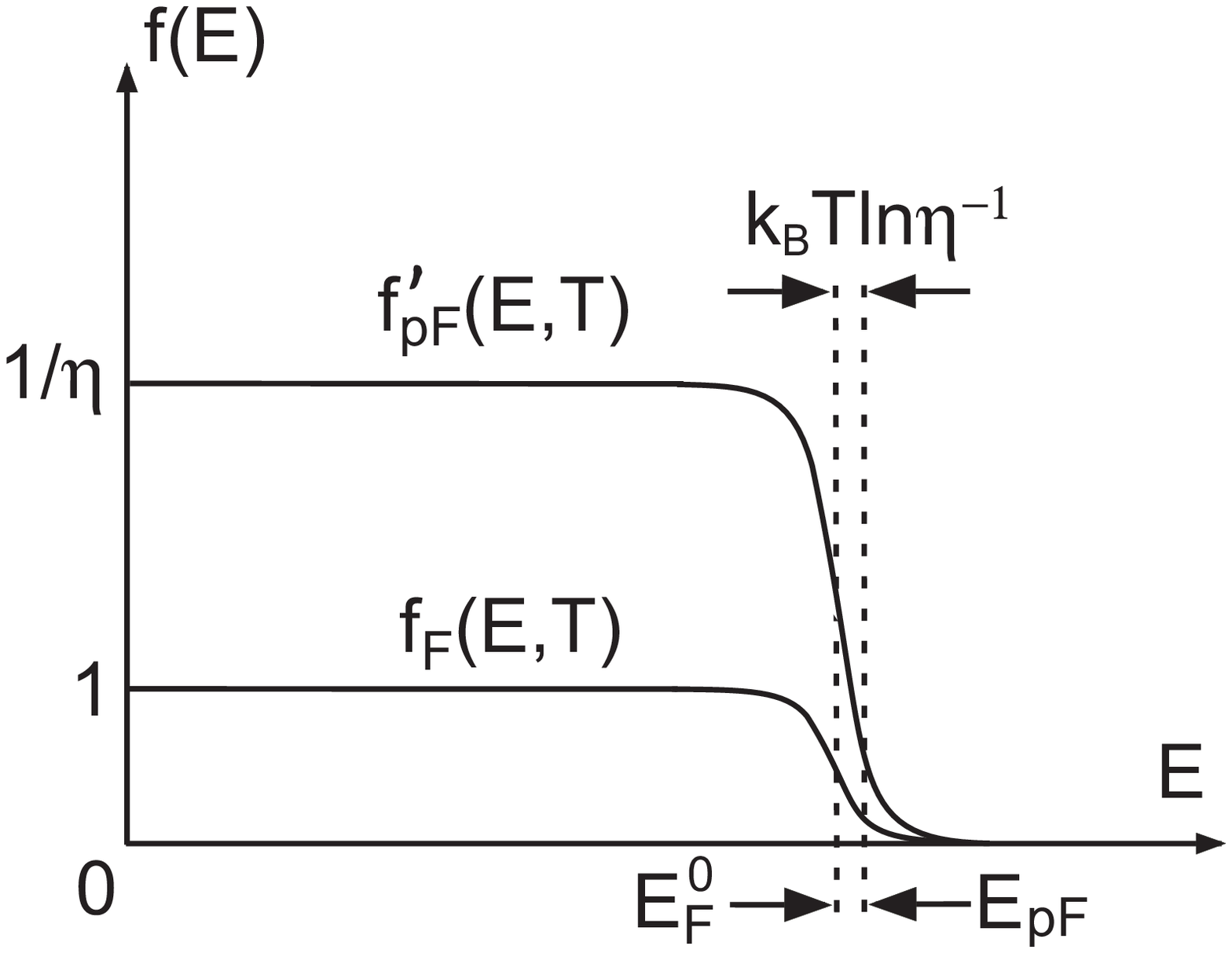}
\caption{\label{fig1}Schematic illustration of the distribution
functions of the Fermi statistics $f_{F}(E,T)$ and partial Fermi
statistics $f'_{pF}(E,T)$ with the same total density of states.
The two dashed lines indicate the zero temperature Fermi energy
$E_{F}^{0}$ and partial Fermi energy $E_{pF}$, respectively. Their
difference is about $k_{B}T ln \eta^{-1}$.}
\end{figure}

\begin{figure}
\includegraphics{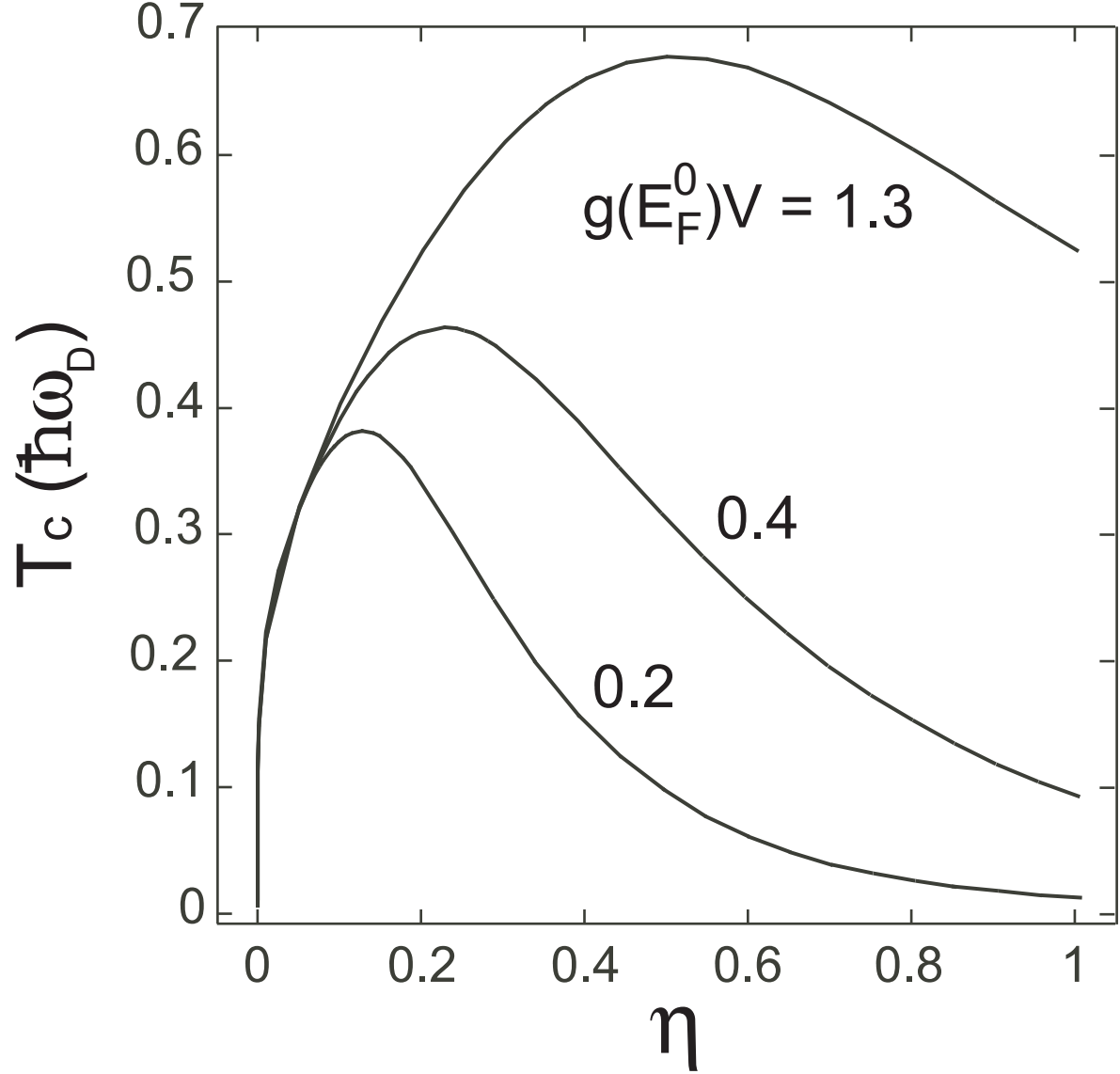}
\caption{\label{fig2}Calculated $T_{c}$ of partially localized
electron systems as a function of $\eta$. The calculations are
based on the partial Fermi statistics and a BCS-type interaction
between electrons. The unit of $T_{c}$ is $\hbar\omega_{D}$. Three
different curves correspond to three different values of
$g(E_{F}^{0})V$ with $g(E_{F}^{0})V$ = 0.2, 0.4, and 1.3,
respectively.}
\end{figure}

\end{document}